# Study on Transient Spectrum Based on Charge Transfer Of Semiconductor Quantum Dots

Zhexu Xi[1], Hui Zhao[2]

Bristol Centre for Functional Nanomaterials, University of Bristol[1]

College of Materials, Xiamen Univeristy[2]

Abstract: With the increasing energy crisis and the prevalent concept of green sustainability, quantum dot materials have become a hot spot in the academic and industrial fields of chemistry. Due to unique, tailor-made photovoltaic properties based on marked quantum-confined effects, it's necessary to identify the QD-based charge transfer process connected with a lifetime of stimulated excitons. Additionally, inorganic nanoparticles with a continuum of electron states contribute to the consistency between electron dynamics and their function through complexation with QDs. Ultrafast spectroscopy can be widely used in this system, the most typical of which is the time-resolved transient absorption spectroscopy, especially on a femtosecond or picosecond timescale. In this paper, we used the ZnSe/CdS core-shell quantum dot as the donor, and the $TiO_2$ film as the metal oxide molecule as the acceptor, through steady-state and transient absorption techniques. Within, the electron transfer and related processes between the two composite systems were explored, and the relationship between the electron transfer rate constant ($k_{BET}$) and particle size and QD core size was further studied. Through the research content of this paper, it is hoped to provide materials for quantum dot sensitization devices with more controllable features.

1. INTRODUCTION

   A quantum dot is a "zero-dimensional nanomaterial" whose size is smaller or closer to the Bohr radius. The quantum confinement effect is particularly remarkable due to the limited conduction of conduction band electrons, valence band holes and excitons in three-dimensional space. From the energy point of view, quantum dots have energy levels of splitting discontinuities, and energy band structures have strong size dependence. From a functional point of view, quantum dots have adjustable band gap, high photoluminescence quantum yield, and narrow luminescence peaks. Compared with traditional organic dyes, quantum dots have high-quality size-dependent optical properties and photochemical stability. Therefore, quantum dots have wide applications in bio-fluorescence imaging, next-generation photovoltaic devices, sensors, and optoelectronic displays. In addition, semiconductor quantum dot sensitization systems and devices introduce more metal oxide nanoparticles as part of the coupling structure, which is the result of scientific research and industrial creation that exhibits high-quality properties based on its electronic structure, namely its particles. The dimensionally sensitive electronic structure is easy to develop a corresponding functional device based on tailor-made electronic characteristics, showing a strong development potential.

2. QUANTUM DOT OVERVIEW

2.1 Basic Concepts And Types Of Quantum Dots

   Nanocrystals are hundreds of thousands of atomic polymers ranging in size from 1 to 100 nm. Their microscopic quantum features ensure the unique electronic band structure of nanomaterials. Through the coupling between microscopic atoms, the electronic structure or properties of nanomaterials can be regulated without changing components or excessive molecular processing, thereby generating new functions, such as highly integrated intelligent storage functions of self-assembled carbon nanotubes, and the superparamagnetism of nanomagnetic materials and the size dependence of the band gap of semiconductor nanocrystals [1].

   The quantum dot is a nanocrystalline semiconductor structure in a zero-dimensional system. In the classical sense, its three-dimensional size is between 1-10 nm, that is, less than or close to the exciton Bohr radius. Since the exciton motion in quantum dots is limited in three dimensions, under the zero-dimensional structure, the quantum dot density function exhibits a hydrogen-like spectral discrete line with the change of energy, replacing the continuous energy band of the bulk material structure [2].

   Common quantum dots are mainly divided into three categories: they can be composed of a single semiconductor, such as the II-VI family (CdS, CdSe, ZnSe, etc.), the III-V family (InP, InAs, etc.), and the perovskite family ($CsPbX_3$, etc., wherein X is a halogen), a monovalent (Si, etc.), an IV-VI family (PbS, etc.), and an I-III-VI family ($CuInS_2$, etc.); The materials are composed of a core-shell structure, wherein the binary core-shell structure mainly comprises two types: a wide band gap semiconductor core-narrow band gap semiconductor shell and a narrow band gap semiconductor core-wide band gap semiconductor shell [3].

   The II-VI family of quantum dots has the highest quantum efficiency and excellent luminescence stability. The excitation spectrum is narrow, symmetrical and finely tunable in the visible region, but most of them are highly toxic. The III-V family of quantum dot materials is less toxic, but optical properties is unsatisfactory. Perovskite quantum dot materials have become the focus of research because of simple synthesis methods, low cost and high luminous yield. For example, the half-peak width of the excitation wavelength is narrow, and the quantum efficiency is over 90%, but its stability is still a urgently problem needed to be solved[4].

## 2.2 Physical Effects Of Quantum Dots

Quantum dots, due to their quantum size and extremely small specific surface area, make their physical and chemical properties very different from those of conventional systems. With the innovation of device processing and performance optimization engineering, the performance of quantum dots is gradually in LEDs. Photoelectric detectors, crystal field effect transistors, bio-imaging, solar cells and other fields have received extensive attention and discussion. The quantum effects it exhibits are as follows:

### 2.2.1 Quantum size effect

Due to the high specific surface area of the quantum dots, the band gap energy between the highest filling track and the lowest unfilled track in the molecular bonding orbital can be adjusted by adjusting the particle size and composition of the quantum dots, thereby controlling the binding energy of the excitons. As the particle size decreases, the energy level structure changes from a quasi-continuous structure to a discrete state, the electron and hole wave functions overlap, and the band gap becomes wider. The composite excitation spectrum is blue-shifted [5]. This process causes the quantum dot particles to have a nonlinear optical response with higher optical performance, and accordingly, this property is also a direct manifestation of the quantum dot material distinguished from the general bulk material.

### 2.2.2 Quantum confinement effect

When the quantum dot size is close to the Bohr radius of its semiconductor material, the binding energy of the electron in three-dimensional space is increased, so that the mean free path of electron motion is reduced, and the localization of the electronic state is deepened [6]. When the mean free path of electrons is sufficient compared with the size of the quantum dot system, the degree of coherence of the electron wave function with the same vibration frequency and constant phase difference between the nano-interfaces is enhanced [7]. The electrons and the hole wave function are more overlapped, and the electrons are more likely to bind to the holes to form an exciton pair, and the probability of exciton generation increases. Therefore, by this effect, the radius of the quantum dots can be changed to gradually increase the energy level interval to form a discrete shape, thereby realizing the regulation of the band gap width [8].

The enhancement of the dielectric properties of quantum dot systems leads to the enhancement of the properties of the localized domains mainly due to the enhancement of surface and internal local field strength.

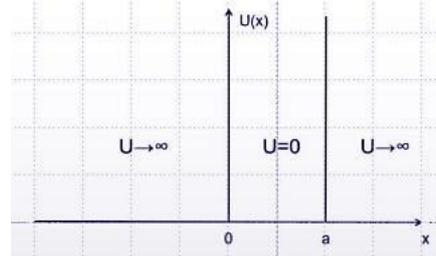

Figure1 Schematic diagram of one-dimensional infinite depth and potential well model

Applying this effect to the energy band width adjustment model of quantum dots, taking the simplest one-dimensional wireless body well as an example, the relationship between the energy eigenvalue and the quantum number n is easily obtained by quantum mechanics:

$$E_n = \frac{\pi^2 h^2}{2ma^2} n^2 \quad (1\text{-}1)$$

Assuming that the system consists of n particles and the interparticle interaction is negligible, the energy band width $E_g$ of the system is:

$$E_g = E_{n+1} - E_n = \frac{\pi^2 h^2}{2ma^2}(2n+1) \quad (1\text{-}2)$$

It is known from the formula that the smaller the potential well, the larger the band gap, which can explain the smaller the quantum dot size and the blue shift of the exciton transition wavelength.

### 2.2.3 Quantum surface effects

As the particle size of quantum dots decreases, the number of defects occurring on the surface increases. The specific surface area increases, resulting in insufficient coordination of lattice atoms, resulting in surface defects caused by unsaturated bonds and dangling bonds, or lattice defects and impurities produce defect states [9]. These surface defects are often used as new non-radiative recombination centers, which can reduce the lifetime of carriers, increase the surface energy, and change the carrier transport and configuration near the surface of quantum dots [10]. The generated surface trap electrons or trap holes can further hinder the photoelectron trapping ability of the quantum dots and affect the macroscopic physical properties such as the luminescent properties of the quantum dot material. Surface effects can make semiconductor quantum dots have greater surface energy and higher surface activity.

### 2.2.4 Quantum tunneling effect

When the total energy of a quantum dot nanoparticle system is lower than the barrier, from the perspective of classical mechanics, electrons are not enough to cross the energy level barrier, while in the quantum mechanical system, when electrons enter the region where the total energy of the particles is lower than the potential energy, there is a certain probability running through this area, called quantum tunneling [11]. Taking a simple one-dimensional potential box as an example, according to the theory of quantum mechanics, the quantum has volatility, and the tunneling behavior satisfies the Schrödinger equation. The solution of the equation represents the probability density of particles appearing in various regions. The probability of a particle crossing a barrier is defined as the ratio of the probability density of the transmitted wave to the probability density of the incident wave:

$$D = \frac{|\varphi_{III}(a)|^2}{|\varphi_I(0)|^2} \xrightarrow{\text{According to boundary conditions}} \frac{|\varphi_{II}(a)|^2}{|\varphi_{II}(0)|^2} \quad (1\text{-}3)$$

And the lower the barrier height $V_0$, the smaller the barrier width α, the greater the probability of particle tunneling [12].

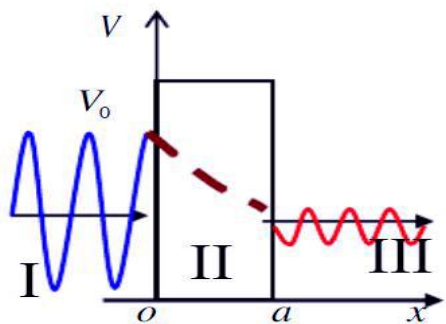

Figure 2 Schematic diagram of quantum tunneling effect of one-dimensional potential box

Scanning tunneling microscopy technology utilizes this effect to overcome the wavelength limitation and aberration limitation of traditional microscopes by tunneling electrons to scan the surface of the object and then presenting the topographic features of small-sized objects at high resolution [13].

In addition, this phenomenon of special relativity is widely used in micro-nano-scale device processing [14][15]. In the FET device, an ultra-thin tunneling insulating layer (such as h-BN) with an atomic thickness is introduced into the interlayer of a single-layer transition metal disulfide ($MoS_2$, $WS_2$, etc.), which is beneficial to improve the interface high Schottky barrier height to achieve high output current and high carrier mobility at room temperature.

2.2.5 Coulomb blocking effect

Due to the nano-scale size of quantum dots, the electron spacing is small. Due to the quantum confinement effect, the motion of electrons is strongly bound in three-dimensional space. Therefore, as the space is stronger, the coulomb rejection between electrons is greater, that is, When the first electron enters the quantum dot, the increased electrostatic potential of the system is greater than the thermal motion of the electron, hindering the second electron and failing to proceed. This effect can visually explain the tunneling effect of a single electron, that is, since the Coulomb island is small enough in size and electrically insulated from the outside, its capacitance to the outside is small enough as $10^{-16}$ F, therefore, electrons must overcome the Coulomb repulsion of another electron with sufficient energy when tunneling [16]. At the same time, in application, the single-electron functional device uses the single-electron tunneling effect in the nano-tunnel junction to achieve low power consumption, very small size device function, and the smaller the device size, the better the performance [17].

2.3 Optical Properties Of Quantum Dots

Based on a series of physical effects at the above quantum scale, in addition to high fluorescence intensity and easy to achieve high resolution detection, quantum dots produce many unique optical properties. First, due to the significant quantum size effect, quantum dots cause the energy level structure to exhibit a discontinuous discrete structure different from the bulk material. As the particle size decreases, the band gap becomes wider, and the electron-hole complex excitation spectrum undergoes blue shift. And by adjusting the size and composition of the quantum dots to finely adjust its emission spectrum, it is excited to a different color. Secondly, quantum dots as a class of inorganic materials have good photochemical stability compared with traditional organic dyes, which can be slow in long-time illumination and sensitive to water oxygen in the environment [18]. According to this principle, it can be used as a marker in living organisms, anti-photobleaching, suitable for long-term stable excitation dynamic observation and result archiving, as shown in the following figure.

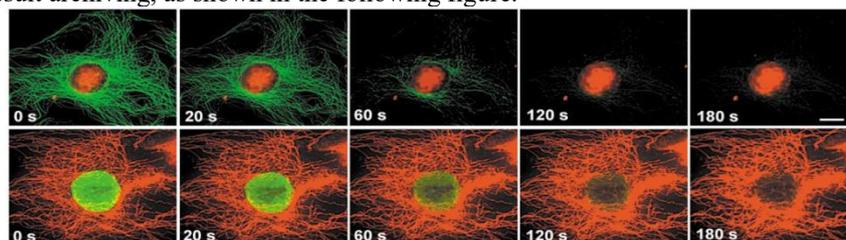

Figure 3 compares the resistance of QD608 to the fluorescent dye Alexa 488 for photobleaching and multicolor labeling [18]

(The nuclear antigen in the first row is labeled with the red color of the QD608-chain enzyme antibiotic protein, the microtube is marked with the green color of Alexa 488, and the color chart of the reverse operation in the second row is used to record the color decay during the continuous exposure time (s).)

Thirdly, due to the uniformity of the morphology of the quantum dot grains and the good dispersibility after solution treatment, the emission spectrum is narrow and symmetrical, similar to the Gaussian function image, which is beneficial to reduce the quantum dots of different sizes during the multi-color excitation process. At the same time, its excitation spectrum is wide, which is beneficial for marking and simultaneously exciting lasers of different wavelengths. The two advantages combine to facilitate the formation of complex white light, so the purity of the quantum dot luminescence is higher.

Fourth, the second most significant feature of organic dye molecules is the larger Stokes shift, which is likely to reach 300-400 nm [19]. It can reduce the overlapping area of the excitation spectrum and the emission spectrum, which is beneficial to eliminate the background interference of the spectral signal and improve the sensitivity.

Fifth, the fluorescence lifetime of quantum dots is relatively long, generally reaching 20-50 ns. The lifetime of Si quantum dots with quasi-continuous band gap can exceed 100 μs [20]. Usually, the fluorescence of most substances has decayed and the fluorescence of quantum dots still remains. It exists and therefore, facilitates obtaining an interference-free optical signal. At the same time, quantum dots have good biocompatibility and are commonly used as fluorescent probes for non-toxic in vivo labeling and detection in vivo.

In addition, from the high performance of quantum dot functional devices, wet-synthesized colloidal core-shell semiconductor quantum dots are the focus of research. Compared to the energy-injection technique, the wet synthesis is more conducive to the adjustment of optical properties and the realization of new functions: it matches the precise morphology and size control, and has high photoluminescence quantum yield and spectral purity. Moreover, the epitaxial vacuum environment is more difficult to achieve than the solution-based dispersibility and is therefore more expensive to operate [21]. In addition, the outer shell coating facilitates passivation of non-radiative binding sites on the surface, enhancing the ability of the shell to limit the excitons.

However, quantum dot materials still have the following significant difficulties in the promotion process:

(1) The resistance to water, oxygen and heat is still insufficient: in the quantum dot material of the core-shell structure, the surface ligand plays an important role in maintaining the stability of the quantum dot structure and maintaining efficient quantum light emission. However, due to adverse factors such as photothermal water and oxygen, the ligand on the surface of the quantum dot may fall off or fail due to chemical reaction. When exposed to light or heat, as the internal energy of the ligand increases, the atomic motion increases, the ligand detaches from the surface of the quantum dot, or the probability of ligand fragmentation increases greatly. Under water and oxygen conditions, complex chemical reactions occur on the surface ligands of quantum dots, which affects the luminescence stability of quantum dot materials. The instability of quantum dots is mainly reflected in three aspects: the fluorescence emission intensity is reduced, the peak position is shifted and the half width is widened. The instability caused by these various factors can greatly affect the application of quantum dots in devices. It is necessary to block water and oxygen, and to ensure the temperature, which puts high requirements on the packaging of the device.

(2) Agglomeration of quantum dot materials: In the process of application, the agglomeration of quantum dots is also an important reason for affecting the luminescence of quantum dots. Since the principle of quantum dot luminescence is a quantum confinement effect, stable luminescence properties can be maintained only when the quantum dot volume is small. When the quantum dots are agglomerated, the size of the quantum dots is changed, thereby invalidating the quenching of the quantum dots.

Therefore, there is an urgent need to solve the problem of how to maintain the relatively stable optical properties of quantum dot materials in the process of application.

3. ELECTRON TRANSFER BEHAVIOR OF PHOTOINDUCED INTERFACE AND MARCUS THEORY

3.1 Marcus Classical Electron Transfer Theory

Rudolph Marcus, a professor at the California Institute of Technology, was awarded the 1992 Nobel Prize in Chemistry for his research on the theoretical model of electron transfer between 1956 and 1965. He believes that the rate of electron transfer process depends on the distance between the electron donor and the acceptor, before and after the transfer. The change in free energy and the size of the recombination energy of the dispersion medium [22]. This model is useful for calculating the energy change, electron transport and its relationship with the surrounding environment when two electrons are close to each other. Therefore, it is more used in the future to explore the electron transfer between semiconductor quantum dots under light induction state dynamics to improve optical performance such as photoelectric conversion efficiency of quantum dot nanodevices.

There are two main types of common electronic reaction pathways: intramolecular transfer and intermolecular transfer. Intramolecular transfer involves the cleavage and recombination of chemical bonds. On the contrary, intermolecular transfer directly undergoes charge separation and recombination through the action of a dispersant such as a solution. The specific process is shown in the following figure:

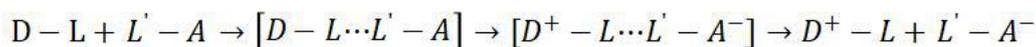

$$D - L + L' - A \rightarrow [D - L \cdots L' - A] \rightarrow [D^+ - L \cdots L' - A^-] \rightarrow D^+ - L + L' - A^-$$

Figure 4 Flow chart of electron transfer reaction between donor (d) and acceptor molecule (a)

According to the transition state theory, intermolecular transfer is the most favorable way of energy. Marcus electron transfer theory is also mainly applied to such transfer pathways. Specifically, the electron transfer process during the transition from the reactant state curve R to the product state curve P, the electron donor and the acceptor form the precursor D|A, and then pass through the energy barrier at a to reach the transition state B, and then relax to the low-energy stable product C, and then form the subsequent network. The state of $D^+|A^-$, where the reactants are close to the product energy at B, can undergo electron transfer process [23].

In the classic Marcus model, the following basic preconditions must be met: when the electrons move from the reactant state to the product state during the electron transfer process, the Blanck-Kanten principle states that the electron transfer process between the molecular vibrational levels is caused by the vibrational energy level wave. The degree of overlap coupling of the functions is determined; at the same time, the energy conservation, the energy of the regions before and after the electron transfer must be equal; the matrix elements of the electron coupling should be large enough to allow the reaction to be completed under adiabatic conditions.

Under this premise, the electron coupling effect in the electron transfer process is very strong. Therefore, the nuclear vibration at a small frequency under adiabatic approximation can be neglected relative to the high-frequency electronic vibration, and the classical Marcus electron transfer theory model is obtained.

If we assume that the vibration of electrons in the transmission direction between systems is a simple harmonic vibration, Marcus et al. reasoned that the activation energy change before and after electron transfer is related to the free energy change:

$$\Delta G^{\neq} = \frac{(\lambda + \Delta G)^2}{4\lambda} \quad (1\text{-}4)$$

From (1-2), it can be concluded that in the classical et theory, it can be divided into three cases according to the relationship between the system reorganization energy and the reaction driving force before and after electron transfer:

(1) When the reaction driving force is less than the recombination energy, the heat released by the reaction is larger, and the reaction rate is larger;

(2) When the reaction driving force is close to the recombination energy, the reaction has no energy barrier and the rate reaches the maximum;

(3) When the reaction driving force is far away from the recombination energy, an inversion zone is generated, and the larger the heat released by the reaction, the smaller the reaction rate.

According to the above theory and formula, the classic Marcus electron transfer formula can be obtained:

$$k_{ET} = \frac{1}{2\pi}\sqrt{\frac{f}{\mu}} \exp\left[-\frac{(\lambda + \Delta G)^2}{4\lambda k_B T}\right] \quad (1\text{-}5)$$

Where $k_{ET}$ is the apparent rate constant of the electron transfer process, h is the Planck constant, and $k_B$ is the Boltzmann constant. $\lambda$ is the restructuring of the system, $\mu$ is the study of the quality of the particles in the system, $\Delta G$ is the Gibbs free energy change before and after electron transfer.

3.2 Analysis Of Related Physical Quantities Of Electron Transfer Behavior

Through the above formula and analysis, the influencing factors of electron transfer behavior mainly include $\overline{H}(E)$、$\lambda$、$\Delta G$ three physical quantities, the meaning of which is the coupling function before and after electron transfer, the reorganization energy of the reaction system and the Gibbs free energy variation of the electron transfer process.

$|\overline{H}(E)|^2$ is the degree of coupling overlap of the electron orbits in the initial state of the electron transfer process, which is the coupling integral of the 1s electron orbit and the lowest unfilled orbit without considering the thermal electron transfer. $|\overline{H}(E)|^2$ is also related to the distance between the electron donor and the receptor [24].

$\lambda$ for the reorganization of the system, when the potential energy surface of the reactant is close to the potential energy surface of the product, the internal energy level structure of the electron donor and the acceptor and the orientation arrangement of the surrounding solvent molecules will be reformed. There is a correlation with the rate constant of the electron transfer reaction. The contribution of recombination energy mainly comes from two parts: internal domain reorganization energy and peripheral reorganization energy. The internal recombination energy mainly indicates the internal structure change of the donor and the receptor in a balanced conformation before and after electron transfer, and the external recombination energy is the solvent recombination energy, that is, the energy generated by the directional polarization or orientation of the external solvent molecules.

$\Delta G$ is the amount of Gibbs free energy change before and after the electron transfer reaction, and -$\Delta G$ can be used to indicate the driving force of the system reaction. Under the research system, the change is mainly affected by three factors: $\Delta G_{charge}$ represents the freedom of charge energy refers to the difference in energy imparted or received by a positively or negatively charged substance; $\Delta G_{coulomb}$ represents the energy required for coulomb free energy, ie, the separation of space electron-hole pairs; $\Delta G_{electron}$ represents the free energy of electrons, ie the energy difference of the state electron transfer. Only the free energy of the electron can be measured:

$$\Delta G_{electron} = E_{electron\,acceptor} - E_{1Se} \quad (1\text{-}6)$$

The total free energy of the system is expressed as:

$$\Delta G = E_{electron\,acceptor} - E_{1Se} + \frac{e^2}{2R_{QD}} + 2.2\frac{e^2}{\varepsilon_{QD}R_{QD}} - \frac{e^2}{4(R_{QD}+d)}\frac{\varepsilon_{electron\,acceptor}-1}{\varepsilon_{electron\,acceptor}+1} \quad (1\text{-}7)$$

among them $E_{electron\,acceptor}$ with $E_{1Se}$ are the conduction band bottom energy of the acceptor and the donor (quantum dot), respectively. $R_{QD}$ is the radius of the quantum dot, $\varepsilon$ is the dielectric constant of each, and d is the distance between the donor and the acceptor.

4. TIME-RESOLVED TRANSIENT ABSORPTION SPECTROSCOPY

4.1 Steady-State Absorption Spectroscopy And Pump-Detection Techniques

All matter is selective for the absorption of light. A molecule or atom in the ground state absorbs light of a specific wavelength range, absorbs the energy of the photon, and thus transitions to a higher energy level, that is, an excited state level, to generate a spectrum. By measuring the absorption spectrum, the internal energy level structure and composition of the molecule can be ascertained. However, the spectrum in a steady-state environment cannot measure the change in the continuity of excitons, that is, the number of excitons and the change in microscopic state in a period of Δt [25].

The absorption spectrum follows the Lambert-Beer law and describes the relationship between the absorbance A of the substance and the concentration c of the solution to be tested:

$$A = \lg\left(\frac{I_0}{I}\right) = \varepsilon bc \quad (1\text{-}8)$$

Where A is the absorbance, and $I_0$ and I are the incident and transmitted light, respectively. $\varepsilon$ is the proportional coefficient, ie the molar absorptivity, which is related to the wavelength of the incident light and the nature of the absorbing material, c is the molar concentration, and b is the thickness of the absorbing layer, usually in cm.

In order to further understand the evolution of the excitons of particles in the excited state with time, ultrafast spectroscopy technology has been widely concerned. Due to the short pulse advantage of ultrafast pulses, it can interact with multiple degrees of freedom such as charge, lattice, spin, orbital angular momentum, since ultrafast light can interact with these degrees of freedom in femtosecond to nanosecond time. The role, therefore, can be used to sense, detect, and interpret transient interactions between the quantum excited states of matter within the material. This technique has the main features of time-scale resolution and coherent states [25].

Pump-detection technology is one of the most popular categories in ultrafast spectroscopy. The basic principle is shown in the figure:

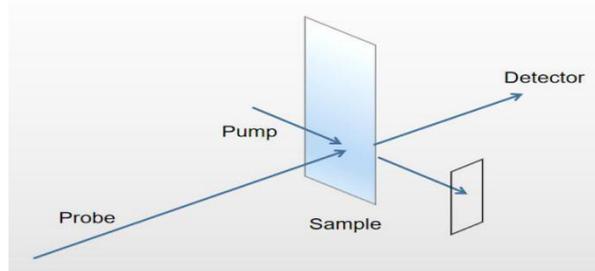

Figure 5 Schematic diagram of pump-detection technology

The ultrashort laser pulse sequence emitted by the optical parametric amplifier in the ultrafast laser system is split into two optical paths via a splitter, and the first laser is pumped to excite the molecules of the substance to be tested for a duration much shorter than the interior of the molecule. A kinetic process of excitation, after a delay time, emits a second laser beam, which is a probe light for detecting changes in the internal structure and related excitons. The probe light is weak, uniform, continuous with respect to the pump light intensity. In the process of transitioning from the excited state of the molecule to the ground state, adjusting the value of different delay times can be achieved by adjusting the optical path difference between the two optical paths.

If the delay time is changed and the wavelength of the excitation light is constant, continuous monitoring can be performed to obtain a distribution signal of the spectrum at the same wavelength as a function of the delay time. If the delay time of the excitation source is changed, the spectrum can be obtained with the same delay time. The distribution signal of wavelength changes. Therefore, during the whole dynamic process detection, the absorption spectrum can be controlled according to different variables to obtain some different characteristic signal peaks.

4.2 The Way Transient Absorption Works

The time-resolved transient absorption spectrum has both the universality of the steady-state absorption spectrum and the time-resolved femtosecond scale. By adjusting the time interval between the pump light pulse and the probe light pulse reaching the sample, under the condition of the delay time of different probe light pulses with respect to the pump light pulse, the change of the light intensity after the probe light passes through the sample is recorded, thereby studying the law of the optical parameter of the excited sample as a function of delay time.

During the experiment, a high-energy pump light is used to excite the sample in the ground state to the excited state, and then a low-energy probe light is used to detect the population of the excited state level of the excited sample. The delay time of the probe light pulse relative to the pump light pulse is obtained according to the change of the particle population on the excited state energy level with the delay time, and the detailed process of the material molecule from the high energy level excited state to the low energy ground state is obtained [26].

Using time-resolved transient absorption spectra to change the delay time each time $\tau$, the measured light intensity $I_{pump}$ excited by the pump light and the transmitted light of the reference pumped light $I_{nopump}$ can be obtained for each time difference. At the same time, the amount of change in absorbance with or without excitation will be captured by the computer. The amount of change is:

$$\Delta A = A_{pump} - A_{nopump} \quad (1\text{-}8)$$

Under the condition that the incident light intensity $I_0$ is known, the instrument will record all delay times according to Rab-Beer's law:

$$\Delta A(\lambda,\tau) = \lg\left(\frac{I_0}{I_{pump}}\right) - \lg\left(\frac{I_0}{I_{pump}}\right) = \lg\left(\frac{I_{nopump}(\lambda,\tau)}{I_{pump}(\lambda,\tau)}\right)$$

（1-9）

### 4.3 Common Signal Peaks For Transient Absorption

There are four kinds of signal peaks in the common femtosecond time-resolved transient absorption spectrum [26]:

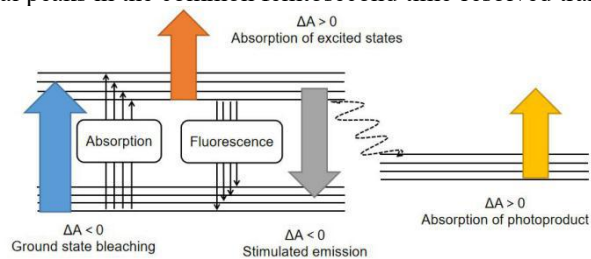

Figure 6 Four signal peaks common to transient absorption spectra

(1) Ground state bleaching (GSB) signal: Under the excitation of pump light radiation, some molecules transition from the ground state to the excited state, so that the number of particles that are originally in the ground state and not excited is reduced, so the ground state absorption of the excited state sample is weaker than that of the light. The excited sample molecules, thus producing a negative signal peak.

(2) Absorption of excited states (AES) signal: a small number of molecules that have been excited by the pump light to the excited state are again excited to the higher excited state level by the excitation of the probe light, so the transmitted light passing through the sample is attenuated, thereby presenting a positive signal peak.

(3) stimulated radiation (SE) signal of excited states: a sample molecule that has been excited by pump light and transitions to an excited state is subjected to the probe light, and stimulated radiation returns to the ground state due to excitation of light of a certain wavelength, and the signal intensity is generally higher than spontaneous radiation is larger and presents a negative signal peak.

(4) Photoproduct absorption signal: a process in which a sample molecule that has been excited by a pump light and transitions to a high-energy excited state returns to the ground state by a series of normal relaxation processes, and then absorbs the probe light to excite the transition to a high-energy excited state signal peak.

### 4.4 Research Significance Of Transient Absorption Spectroscopy

It can be seen from the above that by changing the delay time between the pumping light and the detecting light, the ground state process of the high-level excited state to the low-level thermal vibration of the molecules at different times can be obtained under the condition that the laser wavelength is constant. The transient dynamics of excitons are summarized as two points: First, by detecting the transmitted light intensity at different times, the absorption of transient intermediate products to the optical radiation is obtained, and a series of spectral peak changes are detected. The second is to track the situation of intermediate products [27].

## 5. PHOTOVOLTAIC DEVICES AND QUANTUM DOT SENSITIZED SOLAR CELLS

In the process of vigorously advancing photovoltaic devices, quantum dot devices are undoubtedly the hot spot because the quantum dot size-sensitive electronic energy level structure can be customized to control high performance molecular devices. At the same time, the introduction of the core-shell structure increases the confinement effect of the excitons in the shell, and the non-radiative bonding effect caused by the trap formed by defects or surface states on the passivation surface enhances the photoelectric efficiency, luminous intensity and stability of the quantum dot device. At the same time, the organic ligands connected to the outside of the shell can contribute to the dispersion of the quantum dots in the solvent and the uniformity of the particles, ensuring a high solution modification.

Before introducing a solar cell in which a quantum dot is combined with an inorganic oxide nanoparticle, a sensitizer of a conventional organic dye molecule, usually a ruthenium pyridine compound, is introduced. It acts like a quantum dot and performs charge separation at the interface with the metal oxide.

In order to break through the 32% photoelectric conversion efficiency limit of body cell single-junction solar cells, namely the Shockley-Quezer efficiency limit, the latest research began to introduce metal oxide nanoparticles as electron acceptors and quantum dots to produce nano-sized functional devices [28]. Quantum dots usually function as devices or systems by complexing with other molecules, but because inorganic oxides are continuous oxidation states rather than discrete structures, they can be combined with semiconductors to produce superior performance over single materials: The energy levels of the valence band and the conduction band are staggered, which facilitates the migration of electron holes between the two substances, enhances the lifetime of the carriers, promotes a single process of charge separation or reorganization, and weakens the reverse process, such as $CdS/TiO_2$ system can also broaden the absorption range of the spectral band and improve the capture efficiency of light energy.

The function mechanism of the quantum-dot-based solar cell is explicitly described here. The electron first enters the electron acceptor (MO) from the quantum dot (QD), and the hole is in the QD. After the electron is transported in the MO interface film, it enters the working electrode surface through the external circuit load, and then enters the counting electrode and the electrolyte solution. The oxidative electrolysis reacts, after which the electrons recombine with the holes remaining in the quantum dots and return to the original QD sensitizer.

Ideally, this process cycles back and forth to ensure proper energy conversion of the photovoltaic device. However, due to some reactions that compete with the ideal cycle: the electron scavenging reaction of the electrode to MO, the internal reaction of QD and its hole-electron pair reaction with MO and electrolyte, these reactions cause electron transfer between the receptors. The rate does not match the actual device performance.

# 6. RESEARCH BACKGROUND AND RESEARCH IDEAS

## 6.1 Research Background

Quantum dots have attracted wide attention because of their unique size correlation. the device processing and optimization process is easy to adjust and facilitates excellent functions, while the core-shell structure makes it a sensitizer compared to conventional dye molecules. The solution is more processable and optically stable. At the same time, the wide band gap semiconductor acts as a shell to help passivate the non-radiative binding sites on the surface, promote quantum localization and improve photoelectric conversion efficiency. In addition, it has been found through research that the unique electronic structure properties of quantum dots can be effective by combining them with other substances. By combining with MO nanoparticles that are more stable, less expensive, and more conducive to charge separation, their performance matches the electronic structure to a greater extent. Titanium dioxide, as the most typical type of photochemical reaction and photoinduced effect, is the most widely used transition metal oxide because it is cheap, non-toxic, non-photosensitive, and can absorb a wide range of light. It is focused on its nano-size effect and its regulation of photoelectric performance [29]. $TiO_2$ is not a typical metal oxide film, and the most significant difference from the traditional organic electron acceptor is its continuous state energy level structure. This material acts as a receptor and quantum dot junction, which can be used as an external local level to improve the coupling optical response and size dependence. It is also easier to adjust the electronic structure to achieve performance optimization or new performance development of the coupled system.

## 6.2 Research Ideas And Significance

In this experiment, the principle and mechanism of electron transfer reaction between quantum dots and acceptors can be systematically and comprehensively examined by changing the size of the qd radius to the electronic structure. The femtosecond time transient spectral absorption technique can be used for continuous delay time. The exciton kinetics and relaxation process of the excited state exciton relaxation to the ground state are deeply explored, in order to further understand the relationship between the photoinduced electron transfer rate and the performance of photovoltaic devices, and broaden the application fields of quantum dot sensitized solar cells.

The research ideas are as follows:

(1) Measurement of a series of ZnSe/CdS quantum dots with different particle size and their electron transfer process with $TiO_2$ nano-thickness films by time-resolved transient absorption spectroscopy, and analysis of 1S excitons and QDs inside quantum dots - Photoelectron transfer mechanism between molecules in the -MO heterojunction system;

(2) By comparing the transient absorption experimental data of the single qd and qd-mo coupling system, the kinetics of the transient process and its special spectral signals (such as 1s exciton bleaching signal and charge transfer signal are analyzed), exploring the effects of coupled heterogeneous effects on the evolution of dynamic processes over time;

(3) By comparing the transient absorption experimental data of quantum-doped coupling molecules of different sizes, the kinetics of the transient process and its special spectral signals (such as 1S exciton bleaching signal and charge transfer signal) are analyzed. Investigate the influence of the size of the quantum dots on the evolution of the dynamic process over time in the coupled system, and further understand the limitations of the traditional Marcus charge transfer theory.

# 7. PHOTOELECTRON TRANSFER KINETICS FROM ZNSE/CDS QUANTUM DOTS TO $TIO_2$ FILMS

The core-shell structure quantum dots are widely used in lasers and light-emitting diodes because of their high photoluminescence quantum yield, good luminescence stability, fine spectral tunability, narrow excitation spectrum and symmetry, wide emission spectrum and unique quantum effects, bio-imaging, optoelectronic displays and other fields. Combined with its good solvent dispersibility and grain uniformity, it is important to understand the evolution of its electron transfer reaction and the transient intermediate process, which is helpful to promote its application in device function realization and optimization. The most typical of these is the highly sized, tunable electronic structure that plays an important role in achieving device performance and maintaining high matching of electronic structures and properties. Titanium dioxide is a kind of transition metal oxide which is typically applied in photoexcitation or photoinduction reaction systems. It has not only strong light absorption effect, high light-inducing activity, environmental friendliness, low toxicity and low cost, and is easy to be widely popularized. The energy band structure of $TiO_2$ continuous energy band and the discrete energy level of quantum dots can further adjust the conduction band and valence band position of the system, which is beneficial to the excitation of photoelectrons and the process of separation and recombination. Therefore, it is necessary to explore the transient dynamics and relaxation process of photoinduced electron transfer in this coupled system.

## 7.1 Experimental Ideas

The time-resolved transient absorption spectra were plotted by controlling the two variables of delay time and detection wavelength to measure the microscopic electron transfer process of a series of ZnSe/CdS quantum dots and $TiO_2$ films with different particle sizes, from ground state bleaching and the charge transfer characteristic signal peak and its dynamic evolution process begin to analyze the photoelectron transfer mechanism in the QD-MO system, and explore the QD size (core size) and quantum dot coupling system for the electron transfer process (two major signal peaks and their effects of related kinetics, while using experimental data to verify the limitations of Marcus theory.

Based on the above ideas, the data obtained from the steady-state and transient absorption spectra are divided into two groups: the quantum dot itself and the quantum dot-metal oxide coupling system. The two major signal peaks (ground state) are introduced through quantum dot data. The basic meaning of bleaching signal and charge transfer signal, and the relationship between its signal peak and its kinetic curve and the size of quantum dot nucleus; the second group of QD-MO system is used to explore the two signal peaks and dynamics under the coupled system.

## 7.2 Experimental Implementation
### 7.2.1 Experimental sample information

There are seven different diameter ZnSe/CdS core-shell quantum dot solids in the laboratory, from small to large (unit: nm): 4.9 ± 0.02, 5.3 ± 0.02, 5.8 ± 0.04, 7.4 ± 0.03, 7.1 ± 0.04, 9.2 ± 0.07, 10.5 ± 0.04; laboratory dried $TiO_2$ film.

### 7.2.2 Experimental instrument and spectral test information

The femtosecond time transient absorption spectrum is used to measure the transient absorption spectrum, and the pump light is kept at a constant frequency pulse. The delay time between the pump light and the probe light is adjusted by changing the optical delay line, and then the two beams are concentrated. In the same region of the sample, the band of the probe light is preselected to be measured in the range of 450 to 750 nm, and then the absorption spectrum is recorded. Keep the delay time of pump light and probe light emission constant, and change the light frequency of the probe light to draw the corresponding absorbance change-wavelength ($\Delta A - \lambda$) curve; keep the frequency of the probe light unchanged, adjust the optical delay line to adjust the delay time, and draw the corresponding absorbance change-delay time ($\Delta A - \Delta t$) curve, the kinetic curve under the characteristic signal.

### 7.2.3 Results and discussion
(1) Single quantum dot system

This section introduces the image meanings of two characteristic signal peaks (ground state bleaching signal and charge transfer signal) and their dynamic images through seven different quantum dot images, and analyzes the particle size of the quantum dot size to the peak position. (Unit: nm), peak intensity and the effect of the kinetic process described by the peak, as well as discussion of surface morphology and process.

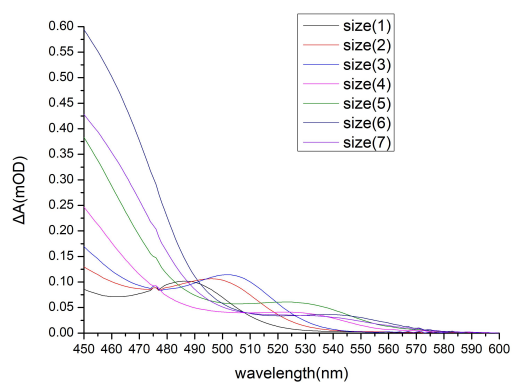

Figure 7 Steady-state absorption spectrum of 7 QD samples

The figure above shows the steady-state absorption spectrum of the QD sample. The minimum wavelength corresponding to the 1S exciton transfer process at each quantum dot size can be seen. For example, the minimum wavelength for achieving exciton transfer in QD sample No. 1 is 485 nm.

According to the position of the signal peak, it can be concluded from the figure that the charge transfer signal in the sample moves to the wavelength decreasing direction as the quantum dot size decreases. The blue shift phenomenon is very obvious because the quantum dot size is reduced. The electrons in the quantum dot are more restrained in the three-dimensional space, and the mean free path is smaller. The energy level gradually changes from quasi-continuous to vertical, and the band gap increases, so the spectrum shifts blue.

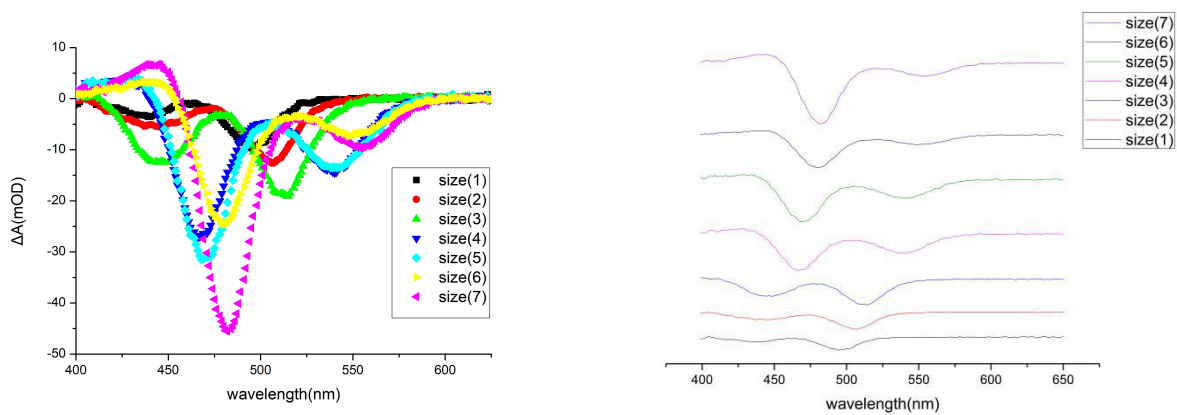

Figure 8 Transient absorption spectra of seven QD samples (left: overall comparison; right: single spectrum, from bottom to bottom, QD samples No. 7-1)

It can be seen from the transient absorption spectrum that the strong signal of the 400-650 nm absorption peak is a series of quantum dot 1S exciton bands due to the state filling of the 1S electron level and the transient intermediate produced by the excitons caused by long-lived bleaching particle composition. Among them, the ground state bleaching signal of the negative absorption caused by the $1S_{3/2}(h)-1S_{1/2}(e)$ interband transition of the quantum dot molecule at a shorter wavelength and the negative absorption at a longer wavelength can be clearly seen. Here, the ground state bleaching means that the quantum dot molecules of a part of the sample are excited by the pump light of a sufficiently large power to cause the ground state molecules to transition to the excited state of the high energy level, so that the number of sample molecules in the ground state is reduced, which is insufficient to absorb enough. The detection of light causes the detection of light to pass through the sample, and thus the transient absorption spectrum exhibits a certain negative signal.

For the position of the signal peak, it can be concluded that the ground state bleaching (at shorter wavelengths) and the charge transfer signal (at longer wavelengths) in the sample all move toward the wavelength decreasing direction as the quantum dot size decreases. The blue shift phenomenon is very obvious. The reason is that the quantum dot size is reduced. The electrons in the quantum dot are more restrained in the three-dimensional space, the mean free path is smaller, and the energy level gradually changes from quasi-continuous to vertical, and the band gap increases. Therefore, the blue shift of the spectrum occurs. The magnitude of the blue shift can be observed carefully. The degree of blue shift is not fixed and the ct peaks of the 6 and 7 quantum dot samples overlap approximately, which may be due to the surface morphology non-uniformity and defect formation. The resulting difference in non-radiative binding sites results.

Secondly, for the intensity of the signal peak:

(1) The ground state bleaching signal decreases with the decrease of the size of the quantum dot. Due to the decrease in size, the specific surface area increases, and the surface adsorption of the quantum dots is enhanced by charge or hole transfer. The sub-life is shortened and the bleaching rate is faster;

(2) The charge transfer signal increases first and then decreases with the decrease of the quantum dot size. First, the quantum dot core size becomes smaller, according to Marcus charge transfer.Theoretically, the electron transfer rate increases as the size decreases.

However, the opposite trend of the quantum size is small enough. The guess may be that the quantum dots with large specific surface area when the wavelengths are close together produce surface localized plasmon resonance, which weakens the peak signal of the resonance electron transfer.

Next, the ground state bleaching kinetics of 1s exciton in quantum dot samples and its correlation with quantum dot size are studied.

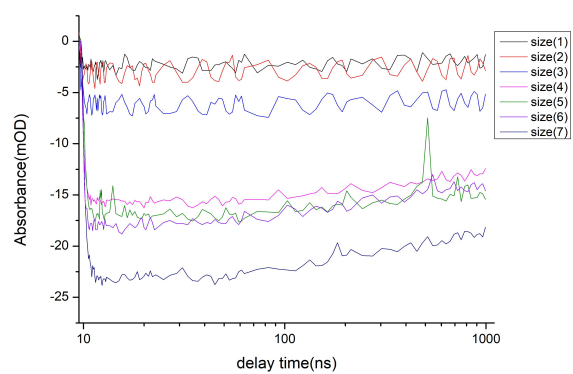

Figure 9 Exciton ground state bleaching kinetics curves for different size qd samples

As shown in the figure, the exciton bleaching process of quantum dot systems of different sizes is divided into the formation of excited state bleach (exciton) and the two processes of ground state bleaching recovery, which correspond to the two segments of steep drop and steady rise in the curve. When the delay time is about 12 ns, the 1S exciton in the quantum dot molecule begins to form. The steep mutation of the curve indicates the formation process of the 1S exciton: the curves of exciton formation under different nuclear sizes are almost identical, indicating that exciton formation is inherent in the molecule itself. At the same time, the larger the nuclear size, the stronger the dynamic process signal, because the larger the size represents the larger surface geometry, the more electrons/holes are involved in the surface, and the more the number of adsorbed molecules; exciton formation afterwards, the change of the steady and moderate rise of the curve indicates that there are few excitons in the empty quantum dot system, and the exciton signal is weakened from the high-level excited state to the low-energy level, which indicates that the exciton lifetime of the single quantum dot system is very long.

(2) Quantum dots - metal oxide coupling system

This section introduces the effects of coupling structures versus single quantum dot systems on the two characteristic signals and their dynamic processes through images of seven different quantum dot-metal oxide coupling systems.

Firstly, the ground state bleaching kinetics of 1s exciton in qd-mo system samples was studied, and the effect of coupling effect on the ground state bleaching kinetics was investigated compared with single quantum dot sample molecules.

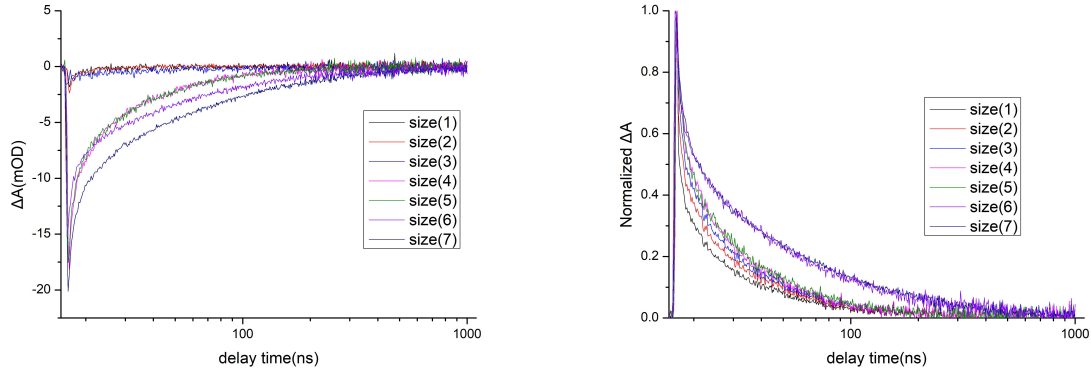

Figure 10 Exciton ground state bleaching kinetic curves for different size QD-MO samples (left: original curve; right: normalized curve)

It can be seen from the figure that the ground state bleaching process is still divided into two processes: 1S exciton formation and bleaching recovery, in which the 1S exciton formation process is unchanged. For the bleach recovery process, it can be seen that the QD-MO system has a significantly higher rate of bleaching quenching or deactivation than the excited state of the single QD system, ie the lifetime of the 1S exciton is greatly reduced (available from No. 1 QD-MO Tens of nanoseconds to 7th QD-MO is close to 1 microsecond). After the heterogeneous system, the $TiO_2$ discrete energy level structure will have stronger coupling with the quantum dot molecules, and the exciton absorption and desorption process completed by the quantum dot surface charge/hole transfer is more intense. As a result of the exacerbation of exciton annihilation, life expectancy is drastically shortened. At the same time, the smaller the quantum dot size, the larger the surface specific surface area, the more excitable surface effects of the excitons and the shorter the exciton lifetime.

Then the kinetics of charge transfer in qd-mo system was studied, and the effect of coupling effect on charge transfer kinetics was investigated compared with single quantum dot sample molecules.

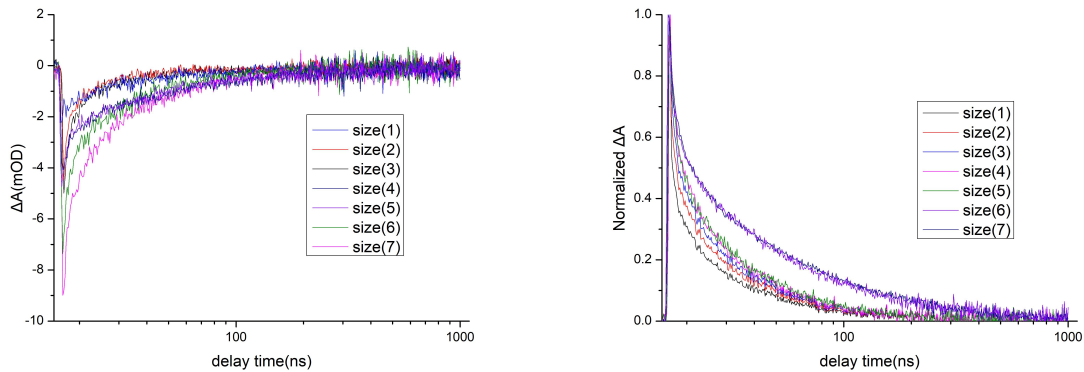

Figure 11 Charge transfer kinetics curves for different size QD-MO samples (left: original curve; right: normalized curve)

It can be seen from the figure that in the charge transfer signal process of the QD-MO system, the smaller the quantum dot size, the steeper the curve is. The reason is that the electron hole transfer process on the surface of the quantum dot is intensified due to the addition of the coupling system of $TiO_2$. The exciton lifetime is shortened, so as the quantum dot size decreases, not only the exciton-induced transition rate between the ground states is larger, but also the charge transfer level recovery rate is larger.

7.3 Summary

Transient spectroscopy studies for QD-MO coupled systems:

(1) Pre-experiment by finding the relationship between the optical signal and the two variables of delay time and detection wavelength to find the best test condition;

(2) Through the same single-size single QD and QD-MO samples, the bleaching signal peaks and electron transfer signal peaks generated by the 1s exciton transition are studied, and the position of the peak signal and the absorption intensity are coupled with the quantum dots. The relationship between systems or quantum dot sensitization devices, and we explored the related kinetic processes, including exciton bleaching and bleaching recovery kinetics and electron transfer kinetics, and studied the kinetic history and its correspondence with quantum dot coupling. ;

(3) Through the 7 different size QD or QD-MO samples, the relationship between the position (wavelength) of the two characteristic peaks and the absorption intensity and the quantum dot size in (2) was studied, and the corresponding dynamics of the two processes were also studied. The process of learning and the variation of particle size and size of quantum dots.

8. CONCLUSION AND OUTLOOK

In this experiment, through the transient absorption spectroscopy study of the photoinduced charge transfer process of quantum dot-metal oxide nanoparticle coupling system, the evolution law of exciton light excitation effect in the core-shell structure with time and the relaxation of corresponding materials are deeply understood. Analysis of band structure, electronic structure and related quantum effects in the process of helium, and based on the introduction of metal oxide and quantum dot coupling to form QD-MO complex system, control nuclear size, etc.

The specific results are as follows:

(1) Graphical analysis of the effect of quantum size and surface factors on the quantum confinement effect on the transient absorption peak and related kinetics;

(2) It is found that the exciton bleaching recovery kinetic curve can be used as a reference standard for measuring the qd-mo electron transfer rate, and the correlation analysis between the bleaching kinetics and the electron transfer kinetic signal is effectively connected together;

(3) The limitations of the classical Marcus charge transfer theory were successfully verified by experimental data, indicating that the electron transfer process is accompanied by the contribution of the Auger electron effect.

As a combination of semiconductor device processing technology, surface technology and quantum science, solid physics and the internal principle mechanism of quantum dot sensitized solar cells, this research has great potential driving value in academia and industry.

First, the quantum effect of size, particle size, and microscopic morphology is used to further fine-tune the performance of the coupled device system, such as photoluminescence quantum yield, luminescence stability, and photoelectric energy conversion efficiency. The old performance is improved, and new performance is developed, has a wide range of applications, such as fluorescent markers and bio-imaging, photoelectric converters, light-emitting displays and diodes, laser technology, etc.;

Second, it can further understand the influencing factors of electron transfer, energy transfer, configuration relaxation and intermediate product formation during material relaxation, such as the refinement of quantum dot structure, the change of metal oxide species and particle size, and the coupling effect. Mechanism adjustment and vacuum conditions or solvent changes to further explore the effects of photoelectric efficiency and microscopic models of electronic separation and recombination;

The third is to use more environmentally friendly green metal atoms to reduce the widespread use of toxic heavy metal Cd in quantum dot systems, such as Si-type Si quantum dots, chalcopyrite ternary compound photovoltaic materials with 1.5eV band gap and low toxicity etc.

Fourth, although relevant scientific research has verified the limitations of the traditional Marcus charge transfer theory, namely in the driving force ( $\Delta G$) away from the recombination-capable region of recombination energy, the electron transfer rate is still sensitive to the change of driving force, contradicting the "reverse zone" in the traditional theory, but it has not been able to interact with electrons and holes in semiconductors and organic molecules. An ideal model for describing photoinduced electron transfer reactions is explored in the action of electron phonons.

Fifth, after further understanding the charge transfer mechanism, it is more reasonable to analyze the dynamic process between the light-excited electron band and the band, and to accurately understand the mismatch between the electron transfer rate and the actual device operating performance, and explore the causes and solutions.